\documentclass{PoS}
\usepackage{enumitem}
\usepackage{mathptm}  
\usepackage{comment}  
\def\pp   {$pp$}
\def\pd   {$pd$}
\def\pA   {$pA$}

\def\PbA {Pb$A$}
\newcommand{\beq}{\begin{eqnarray}}
\newcommand{\eeq}{\end{eqnarray}}
\newcommand{\be}{\begin{eqnarray*}}
\newcommand{\ee}{\end{eqnarray*}}

\newcommand{\ie}{{\it i.e.}}
\newcommand{\eg}{{\it e.g.}}
\newcommand{\etal}{{\it et al.}}

\renewcommand{\L}{\mathcal L}

\newcommand{\glabcms}{\gamma^{\rm lab}_{\rm cms}}
\newcommand{\blabcms}{\beta^{\rm lab}_{\rm cms}}
\newcommand{\dylabcms}{\Delta y^{\rm lab}_{\rm cms}}

\newcommand{\ct}[1]{{Table~\ref{#1}}}

\def\lsim{\raise0.3ex\hbox{$<$\kern-0.75em\raise-1.1ex\hbox{$\sim$}}}
\def\gsim{\raise0.3ex\hbox{$>$\kern-0.75em\raise-1.1ex\hbox{$\sim$}}}

\def\pp   {$pp$}
\def\pA   {$pA$}

\def\beq     {\begin{equation}}
\def\eeq     {\end{equation}}

\usepackage{lineno}

\title{Prospectives for A Fixed-Target ExpeRiment at the LHC: AFTER@LHC}

\ShortTitle{AFTER @ LHC}

\author{\speaker{J.P.~Lansberg}, V.~Chambert, J.P.~Didelez, B.~Genolini, C.~Hadjidakis, C. Lorc\'e, P.~Rosier
        \\
        IPNO, Universit\'e Paris-Sud, CNRS/IN2P3, F-91406, Orsay, France
}

\author{M. Anselmino, R.~Arnaldi, E.~Scomparin\\
        INFN Sez. Torino, Via P. Giuria 1, I-10125, Torino, Italy
        }
\author{S.J.~Brodsky\\
        SLAC National Accelerator Laboratory, Stanford University, Menlo~Park, CA 94025, USA
        }

\author{E.G.~Ferreiro\\
        DFP \& IGFAE, Universidade de Santiago de Compostela, 15782 Santiago de Compostela, Spain
        }

\author{F.~Fleuret\\
        Laboratoire Leprince Ringuet, \'Ecole Polytechnique, CNRS/IN2P3,  91128 Palaiseau, France
        }

\author{A.~Rakotozafindrabe\\
        IRFU/SPhN, CEA Saclay, 91191 Gif-sur-Yvette Cedex, France
        }

\author{I. Schienbein\\
LPSC, Universit\'e Joseph Fourier, CNRS/IN2P3/INPG, F-38026 Grenoble, France 
}

\author{U.I.~Uggerh\o j\\
        Department of Physics and Astronomy, University of Aarhus, Denmark
        }

\abstract{We argue that the concept of a multi-purpose fixed-target experiment
with the proton or lead-ion LHC beams extracted by a bent crystal would offer a number of 
ground-breaking precision-physics opportunities. The multi-TeV LHC
 beams will allow for the most energetic fixed-target experiments ever performed.
The fixed-target mode has the advantage of allowing for high luminosities, spin measurements 
with a polarised target, and access over the full backward rapidity domain --uncharted until now--  up to $x_F \simeq - 1$.
}

\FullConference{36th International Conference on High Energy Physics\\
		 4-11 July 2012\\
		 Melbourne, Australia}

\begin{document}

\section{Introduction}

Since 2010, the Large Hadron Collider (LHC) has been successfully operating  at 3.5 TeV and 4 TeV for 
the proton beams and 1.38 TeV per nucleon for the lead beams. So far, it has delivered an outstanding luminosity 
--beyond expectations-- in spite of its current limited energy (see~\eg~\cite{cern-bulletin}). In a few months, 
the LHC will be shut down for about two years in order to prepare it for an increase in the collision energy close to the
design one, \ie~ 7 TeV per beam.

So far, the LHC proton and lead beams have been used by four large scale collider experiments (ALICE, ATLAS, CMS and LHCb) and 
by three smaller ones, LHCf, MoEDAL and TOTEM~\cite{Heuer:2012zzb}.  
Projects for detector and accelerator upgrades are being presently discussed and the idea of colliding the high
energy LHC beams with an electron beam is being investigated by the LHeC study group~\cite{AbelleiraFernandez:2012cc}. 

In this context, we believe it is well worth advertising the extremely appealing physics opportunities~\cite{Brodsky:2012vg} 
offered by a fixed-target experiment using the LHC beams extracted by a bent crystal. This would provide the LHC complex with 
a high-potential and very cost-effective experiment complementing the existing running or planned projects. 

It is important to keep in mind the important contributions brought by fixed-target experiments to hadron and nuclear physics: 
particle discoveries ($\Omega^-(sss)$, $J/\psi$, $\Upsilon$,...), first evidence for the quark gluon plasma in heavy-ion collisions,
observations of unexpected QCD phenomena in unpolarised and polarised $pp$ collision as well as in $pA$ collisions.
Most of these discoveries bore on the advantages of the fixed-target mode: 
\vspace*{-0.3cm}
\begin{itemize}[leftmargin=0.5cm] \itemsep-1pt
\item[-] outstanding luminosities thanks to the high density of the target, 
\item[-] an unlimited versatility of the target species, 
\item[-] reduced constraints on the kinematics of the studied particles due to the boost 
between the laboratory and the centre-of-mass frames,  
\item[-]  possibilities for in-depth and ultra precise baseline $pp$ and $pA$ studies for heavy-ion physics.
\end{itemize}

In the present case, it is important to note that, for the first time in the history of fixed-target-experiment,   
the released centre-of-mass system (cms) energy will not be a limiting factor, except for the production of the 
top quark. Therefore, it makes  perfect sense to consider the concept of a multi-purpose detector, which would be the first
of a new generation of fixed-target experiments.

\section{A few words on the bent-crystal-extraction mode at the LHC}
\label{sec:lumi}

 The idea of extracting a small fraction of the CERN LHC beam to use it with a fixed-target experiment
has already pushed forward in the early 90's as an alternative to LHCb~\cite{LHCB} to study flavour physics. This 
idea was turned down, mainly because of the fear of a premature degradation of the bent crystal due
to radiation damages. In the 20 years passed then, bent-crystal beam extraction has become a mature technique which has been tested
many times, at SPS~\cite{Arduini:1997kh}, Fermilab~\cite{Asseev:1997yi}, Protvino\cite{Afonin:2012zz}. It has also been 
a key element in the NA48 kaon experiment at CERN~\cite{Doble:1995ym}.

Nowadays, we know that the degradation due to irradiation is about $6\%$ per 
$10^{20}$ particles/cm$^2$ (see \eg~\cite{Baur00}), whereas the irradiation limit
was only expected to be higher than $10^{19}$  particles/cm$^2$ in the 90's.
$10^{20}$ particles/cm$^2$ corresponds to about one year of 
operation for realistic impact parameters and beam sizes at the crystal location.
The crystal has just to be moved less than a millimetre after a year to let 
the beam impact on an intact spot; this can be repeated almost at will.\footnote{In the heavy ion case, 
 lead-beam collimation/extraction at the SPS has been tested through a 50 mm bent crystal~\cite{Arduini:1997nb}
already more than 15 years ago. 
This showed the feasibility of a large-angle deflection. Further recent tests with a shorter crystal --2mm-- 
to test small-angle deflection have also been performed successfully~\cite{Scandale:2011zz}.
Furthermore, laser ablation techniques have been shown to be successful~\cite{Balling:2009zz} 
in bending diamond crystal, which are extremely tolerant to high radiation doses.}

Bent-crystal beam extraction offers an ideal and cost-effective way to obtain a clean and very 
collimated high-energy beam. Such a technique does not alter at all the LHC 
performance~\cite{Uggerhoj:2005xz,Uggerhoj:2005ms,LUA9}. In the coming years, 
the "smart collimator" solution, originally proposed by Valery Biryukov~\cite{Biry03}
will be tested by the CERN LUA9 collaboration. The proposal~\cite{Uggerhoj:2005xz} 
to "replace" the kicker-modules in LHC section IR6 (where the beam is dumped) by a bent crystal
is another interesting  possibility  to be further investigated. The bent crystal 
will give a sufficient kick to the particles in the beam halo such that they would 
overcome the septum  blade and they would be extracted. 

Such a technique would easily extract a significant fraction of the beam loss, with an intensity reaching
$5\times 10^8$ $p^+$s$^{-1}$. On average, this corresponds to the extraction of mini-bunches
of about 15 $p^+$ per bunch per revolution. In practice, this means that any collision pile-up
can safely be avoided by choosing a reasonably long target such that
at most one of these protons interact with the target on average at a time. This is clearly the case
for a 1-cm thick lead target or a 1-m thick solid or liquid hydrogen target. In this case,
yearly integrated (over $10^7$ s) luminosities of the order of an inverse fb are expected 
(see \ct{tab:lumi-pA})\footnote{Luminosities for the 
Pb run can be found in~\cite{Brodsky:2012vg}.}.
Such high luminosities --far above that of RHIC-- open the path for particle and nuclear physics 
studies of $pp$, $pA$, Pb$p$ and Pb$A$ collisions with unprecedented statistics and in principle 
with no limitation on the particle species --except for the top quark.

\begin{table}[!hbt]
\centering\setlength{\arrayrulewidth}{.8pt} 
\renewcommand{\arraystretch}{0.9}\scriptsize
\begin{tabular}{|ccccc|ccccc|}
\hline
Target     & Thickness  & $\rho$        &$A$                     & $\int dt\L$ 
& 
Target      & Thickness   & $\rho$      &$A$                   & $\int dt\L$
\\
&(cm)  & (g cm$^{-3}$) &  & (fb$^{-1}$ yr$^{-1}$)     & 
& (cm) & (g cm$^{-3}$) &     & (fb$^{-1}$ yr$^{-1}$)\\
\hline\hline 
solid H  & 10 & 0.088 & 1    & 2.6 
&
Cu       & 1 & 8.96  & 64   & 0.42  \\
 liquid H & 100 & 0.068 & 1   & 20
&W        & 1 & 19.1  & 185  & 0.31
 \\
liquid D & 100 & 0.16  & 2    & 24
&
Pb       & 1 & 11.35 & 207  & 0.16 \\
 Be       & 1 & 1.85  & 9    & 0.62
&
  &  &  &  &  \\
\hline
\end{tabular}
\caption{Luminosities obtained over a LHC year with an extracted beam of 
$5 \times 10^8$ p$^+$/s for various targets.}
\label{tab:lumi-pA}
\end{table} \vspace*{-.5cm}

\section{Kinematical aspects}
\label{sec:lumi}

The multi-TeV LHC beams --$E_p=7$ TeV and $E_{\rm Pb}=2.76$ TeV per nucleon-- allow for the 
fixed-target experiment at the highest energies ever reached. With the lead beam, \PbA\  and Pb$p$ collisions 
can be investigated at $\sqrt{s_{NN}} \simeq 72\,\mathrm{GeV}$ ($\sqrt{2E_{\rm beam} m_N}$) as well as \pp, \pd\ and \pA\ collisions 
at $\sqrt{s_{NN}} \simeq 115\,\mathrm{GeV}$ with the proton beam.

In the proton case, the cms is boosted by $\glabcms=\sqrt{s}/(2m_p)\simeq 60$ in 
 the laboratory system. The rapidity
shift between both system is $\tanh^{-1} \blabcms\equiv\dylabcms\simeq 4.8$. Typical particles produced in the cms 
central-rapidity region, $y_{\rm cms}\simeq 0$,  are thus highly boosted such that they emerge in the laboratory frame
at an angle of 0.9 degrees with respect to the beam axis. In the lead case, one has
$\glabcms\simeq 38$ and  $\dylabcms\simeq 4.3$.

In the vast majority of the past fixed-target experiments, the high boost of the particles 
produced at $y_{\rm cms}> 0$ --which is the positive counterpart of the much smaller cms energy 
in fixed-target modes-- has been exploited with a forward detector and
usually with a muon absorber. Owing to the large $\sqrt{s}$ at our disposal and modern
detector technologies, it is perfectly reasonable to look at 
particles produced at large angles with rather small longitudinal momenta but with larger $P_T$. This
would provide a first systematic access to the far backward region, namely $x_F \to -1$, where the 
parton content of the target can be investigated up to very large momentum fractions. This is 
of particular relevance given that deuterium targets, polarised target or nuclear target can be used.
This is an essential asset of the present project.

Doing so, it is absolutely conceivable to achieve a very compact detector geometry, which would not require
a very long, neither wide, cavern. To give a first idea, the LHCb detector as it is, with a rapidity coverage roughly within 
$2 \lesssim y_{\rm lab} \lesssim 4.5$ would give access to $-2.5 \lesssim  y_{\rm cms} \lesssim  0$. 
For the production of a particle such as the $\Upsilon$, $y_{\rm cms}\simeq -2.5$ is equivalent to 
$x_F\simeq - M_\Upsilon/\sqrt{s}\times  e^{2.5}\simeq - 1.1 $.  Clearly, this completely uncharted territory is at reach !

\section{Target polarisation}

Whereas the extracted intensity is clearly large enough to reach outstanding luminosities for
hadron reaction studies, it is not as large that it poses constrains on the choice of a polarised target.
The high energy of the beam means a minimum ionisation and a low heating of the target. 
The heating power due to the AFTER beam would be of the order of 50 $\mu\hbox{W}$ for a typical 
1cm thick target. This allows one  to maintain target temperatures as low as 50 mK. Relaxation times 
can last as long as one month in the spin-frozen mode.
 As regards the damages on the target, they typically  arise after an irradiation of $10^{15} p^+ \hbox{cm}^{-2}$~\cite{meyer}, 
\ie~one month of beam in our case.

In the case of AFTER, the available space can be a major constraint. This would restrict the target choice to 
polarisation by continuous {\it Dynamic Nuclear Polarisation} DNP or a HD target 
which both take less space than the frozen-spin 
machinery. Indeed, in the frozen-spin mode, a dilution refrigerator contains the 
target. It has to be moved from the high-field polarising magnet to the beam.
This usually requires to move  bulky equipments. 
On the contrary, the polarisation of a HD target can be performed outside the experimental site. The polarised 
target can then be transported   to the experimental area at high temperature and low field. Then, a rather standard
 cryostat is needed~\cite{didelez}. 

CERN benefits from a long tradition of DNP for various materials such as NH$_3$, Li$_6$D~\cite{berlin}.  
There are still quite a few experts of DNP all around the world. HD target makers are more rare. One can cite two 
groups: one at TJNAF (USA) and the other at RCNP (Japan)~\cite{kohri}. A rich spin program with AFTER 
has been described in~\cite{Brodsky:2012vg}. We are hopeful that it will motivate
our colleagues working on DNP to revisit  the necessary technology~\cite{solem}.

\section{A selection of flagship studies}

In this section, we limit ourselves to a mere list of a selection of flagship studies. A more complete survey of the physics
opportunities with AFTER can be found in~\cite{Brodsky:2012vg}.

\vspace*{-.17cm}
\subsection{ QCD precision measurements and parton studies at large $x$ in $pp$, $pd$ and $pA$ collisions}

With its outstanding luminosity and an access towards low $P_T$, AFTER, if designed as a multi-purpose detector, 
would definitely be a heavy-flavour, quarkonium and prompt-photon {\it observatory}~\cite{Brodsky:2012vg,Lansberg:2012kf} in 
$pp$ and $pA$ collisions. Large quarkonium yields and precise measurements of their correlations would constrain 
the production mechanisms of quarkonia~\cite{review}.
Thanks to the instrumentation of the target-rapidity region, gluon and heavy-quark 
distributions of the proton (see \eg~\cite{Diakonov:2012vb}), the neutron and the nuclei could then 
be accessed at mid and large momentum fractions, $x$. 
In the nuclear case, the physics at $x$ larger than unity  can be accessed. 
Low-$P_T$ scalar and pseudoscalar quarkonium studies can also be used to access
linearly polarised gluons inside unpolarised protons~\cite{Boer:2012bt}.
In $pd$ collisions, quarkonium production 
--along the lines of E866 for $\Upsilon$~\cite{Zhu:2007mja}--  provides rare
information on the momentum distribution of the gluon in the neutron at relatively small scales.

\vspace*{-.17cm}
\subsection{ Precision spin physics with a polarised target}

Benefitting from  the target polarisation~\cite{Lansberg:2012wj}, one can measure single-spin asymmetries (SSAs) 
in a number of reactions  with a precision far greater than RHIC experiments (see \cite{Liu:2012vn} for DY). SSAs
give direct access through the Sivers effect to the correlation between the nucleon spin and 
the transverse momenta, $k_T$, of the parton in the nucleon. The study of such correlations 
are invaluable in the quest to understand the structure of the spin of the nucleons.

\vspace*{-.17cm}
\subsection{Deconfinement studies in Pb$A$ collisions between SPS and RHIC energies}

A number of observables have been proposed to study deconfinement in relativistic heavy-ion collisions.
Among these, one can highlight the $J/\psi$ and $\Upsilon$ suppression, the heavy-flavour and jet quenching or the production 
direct photons. We note that they could easily be accessed with AFTER~\cite{Brodsky:2012vg}.
In particular, one would take advantage of modern detection technology which should also allow 
for thorough investigations of the behaviour of the quarkonium excited states in the hot nuclear matter. In particular, 
we think of  the $\chi_c$ and $\chi_b$ resonances. This can be envisioned, thanks to the boost of the fixed-target mode,  
even in view of the challenges posed by the high-multiplicity environment of $pA$ and Pb$A$ collisions. 
Such investigation  is of paramount importance in the quest for the quarkonium sequential melting in the deconfined matter~\cite{Lansberg:2012kf} .

\vspace*{-.17cm}
\subsection{$W/Z$ production in their threshold region, ultra-peripheral
$pA$/Pb$A$ collisions, etc. }

As mentioned above, the reduced cms energy of the fixed-target mode will not be a limiting factor. For the first time in
a fixed-target set-up, ultra-peripheral $pA$/Pb$A$ collisions would deliver enough cms energy to produce heavy mesons and
hard di-leptons. In this case,  the nucleus target/projectile $A$/Pb 
is effectively becoming a coherent high-energy photon source, up to $E_\gamma \simeq 200$~GeV. 
A great potential of studies is also offered by semi-diffractive processes due
to the absence of pile-up in such a slow extraction mode~\cite{Brodsky:2012vg}.

The high energy of the proton LHC beams and the outstanding luminosities of the fixed-target mode
will also allow us to study for the first time $W$ and $Z$ boson production close to threshold.
Investigations in the threshold region could be very important to understand QCD corrections
in this part of the phase space, particularly for the search for
heavy  gauge-boson partners in many extensions to the standard model 
(see \eg~\cite{Hewett:1988xc}). The threshold
region would indeed be the phase space probed at the LHC for these conjectured $W'/Z'$.

\section{Conclusion}

The multi-TeV proton or heavy ion beams of the LHC can easily be extracted by means of a bent crystal
positioned in the halo of the beam. A  fixed-target experiment using such an extracted beam offers an 
exceptional --and very cost effective-- testing ground for QCD at unprecedented energies 
and momentum transfers.
We have discussed here the luminosities which can be obtained from the mere extraction of a fraction
of the beam loss. These are significantly higher than those to be recorded at RHIC and competitive with those
of the LHC itself. As discussed, the polarisation of the target does not pose any specific
issues. 

Such a project would enrich the LHC program with a number of studies in polarised and unpolarised
$pp$ and $pd$ collisions, ranging from quarkonium studies to single-spin asymmetries in photon-jet 
correlations. Relativistic heavy-ion collisions can also be studied in Pb$A$ collisions 
in an energy range seldom explored thus far.

\end{document}